\documentclass[journal=jacsat,manuscript=article]{achemso}
\usepackage{graphicx}
\usepackage{hyperref}
\usepackage[version=3]{mhchem} % Formula subscripts using \ce{}

\author{Alberto Martín-Pérez}
\affiliation{Department of Precision and Microsystems Engineering, Faculty of Mechanical Engineering, Delft University of Technology, Netherlands}
\email{A.MartinPerez@tudelft.nl}
\author{Peter G. Steeneken}
\affiliation{Department of Precision and Microsystems Engineering, Faculty of Mechanical Engineering, Delft University of Technology, Netherlands}
\author{Farbod Alijani}
\affiliation{Department of Precision and Microsystems Engineering, Faculty of Mechanical Engineering, Delft University of Technology, Netherlands}

\title[An \textsf{achemso} demo]
 {Optomechanical tuning of nonlinearity in graphene resonators}

\abbreviations{}

\begin{document}

\begin{abstract}

Graphene resonators exhibit unique mechanical and electrical properties, making them promising candidates for next-generation sensing applications. Here, we introduce an optomechanical approach that enables active, reversible, in situ tuning of the nonlinear nanomechanical response of graphene resonators. By exploiting photothermal heating and the resulting optomechanical interaction between the mechanical oscillations and the optical cavity formed by a graphene membrane suspended over a reflective substrate, we achieve dynamic control over both the linear restoring force and the nonlinear dynamic response of the resonator. We show that the strength and sign of the Duffing nonlinearity can be continuously tuned through the optical probe power and cavity depth, enabling transitions between hardening and softening behavior as well as near-complete suppression of geometric nonlinearities. Our results establish a versatile platform for on-demand engineering of nonlinear dynamics in graphene resonators and provide a general strategy for active control of the mechanical response of two-dimensional nanoelectromechanical systems.

\end{abstract}

Nanomechanical resonators are excellent tools for probing the nanoworld providing high-sensitivity devices whose versatility has enabled applications spanning from mechanical computation, including logic gates \cite{Tadokoro2021HighlySensitive}, to nanoscale probes for condensed-matter phenomena \cite{bachtold2022mesoscopic}. In this respect, nanoscale sensing is one of the most prominent applications of nanomechanical resonators, which have demonstrated ultrahigh sensitivity in the detection of various physical quantities, including forces\cite{Binnig1986AFM}, temperature\cite{Mertens2003TemperaturePressure}, pressure\cite{ 10.1063/1.4889744, MARTINPEREZ2025117161}, mass\cite{MARTINPEREZ2025117161}, stiffness \cite{Malvar2016MassStiffness} and optical absorption\cite{Kirchhof2023NanomechanicalAbsorption,Chien2018SingleMolecule}. Their exceptional sensitivity has enabled detection limits down to individual molecules \cite{Naik2009SingleMoleculeMS,Chaste2012Yoctogram} and even single electron spins \cite{Rugar2004SingleSpin}. Moreover, the introduction of low-dimensional materials such as graphene or carbon nanotubes as nanomechanical systems \cite{Steeneken2021Dynamics2D, xu2022nanomechanical}  offers the prospect of further enhancing the sensitivity of these devices thanks to the unique features of these materials and their exceptional aspect ratio.

Regardless of the target application, the performance of a nanomechanical resonator relies on precise control of its mechanical stiffness, which governs both its linear and nonlinear dynamics. Several strategies have been developed to tune this response. Geometric design can be used to tailor the cubic stiffness and even reverse its sign, as demonstrated in resonators made of bulk silicon nitride \cite{Li2024StrainEngineering} and graphene \cite{Yousuf2025GrapheneTrampoline}. Electrostatic actuation provides post-fabrication control of the resonance frequency and can also modify the nonlinear response \cite{sajadi2017experimental, samanta2018tuning}, while thermal tuning changes the mechanical tension through thermal expansion \cite{davidovikj2018chip, vsivskins2025nonlinear}. These approaches offer control over individual devices, but they do not by themselves ensure reproducible dynamics across nominally identical resonators. Furthermore, the response depends on material and fabrication parameters, including Young’s modulus, mass density, built-in tension, thickness, and membrane flatness. Variations in these quantities can therefore shift both the linear and nonlinear stiffness and obscure the behavior intended through geometric tuning. For established bulk materials such as silicon and silicon nitride, mature nanofabrication processes provide comparatively tight control over these parameters. The fabrication of resonators based on low-dimensional materials remains less reproducible, however, with substantial device-to-device variations in tension, thickness, and static shape \cite{sarafraz2024quantifying}. Understanding and controlling additional perturbations introduced during measurement and actuation is therefore essential for the reliable implementation of these materials in NEMS.

Nevertheless, even in a scenario in which all fabrication challenges are addressed, low-dimensional material based resonators face an additional reproducibility challenge: transduction techniques, such as electrostatic or optothermal readout, can introduce additional stress, altering the stiffness of the resonator. This transduction-induced stiffness change is the consequence of the energy exchange between the probe and the device\cite{Aguila2022PhotothermalResponsivity, barton2012photothermal}.  Consequently, the reproducibility of these resonators depends not only on fabrication but also on understanding and quantifying transduction-induced perturbations.

Here, we theoretically and experimentally quantify how optical interferometric transduction and actuation modify the linear and nonlinear stiffness of suspended graphene membranes. We show that this optomechanical effect arises from the combined influence of photothermal heating and membrane thermal expansion. Building on this understanding, we develop a model that obtains active optical control of both the linear and cubic stiffness, allowing the membrane response to be switched from nonlinear hardening to nonlinear softening by tuning the incident light.

\begin{figure}
\centering
    \includegraphics[width=0.5\linewidth]{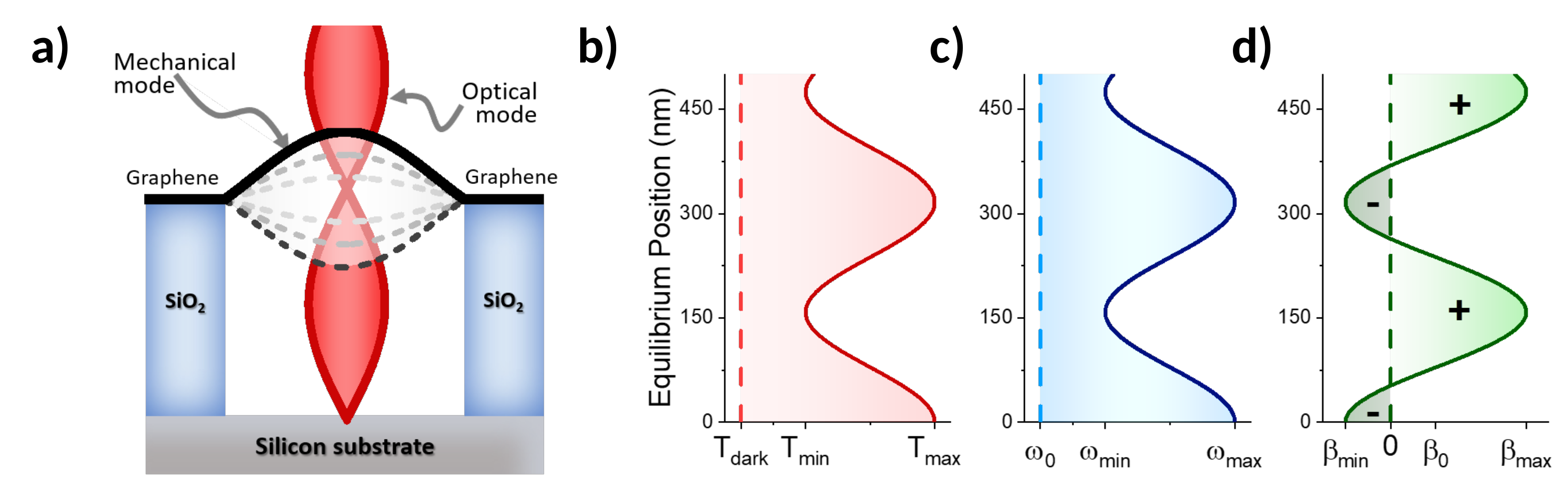}
    \caption{\textbf{Optomechanical interaction. a)} Schematic of the device made of a suspended graphene membrane over a silicon substrate with its optical and mechanical mode. \textbf{b)} Temperature of the membrane calculated as a function of the position. \textbf{c)} Resonance frequency of the membrane calculated as a function of the position. \textbf{d)} Duffing term of the membrane calculated as a function of the equilibrium position }
    \label{fig:fig1}
\end{figure}

Optical interferometry is a widely used scheme for transducing the motion of nanomechanical resonators based on 2D materials. In this approach, a suspended membrane is positioned above a highly reflective substrate so when a laser beam is directed onto it, a standing electromagnetic wave is obtained through the interference of the incident and reflected light fields\cite{Steeneken2021Dynamics2D}, forming an optical cavity (Fig. \ref{fig:fig1}a). The intensity of this standing wave is determined by the membrane-substrate separation (i.e. cavity depth) and the reflectivities of the different interfaces of the device. In this configuration, nanomechanical vibrations of the 2D material modulate the cavity depth, encoding the amplitude of these oscillations in the intensity of the light reflected by the cavity,  which forms the basis of interferometric motion transduction (see Supplementary Information for details).

To investigate how optomechanics affects the membrane motion beyond the linear regime, we model the suspended membrane as a driven Duffing oscillator. Denoting the out-of-plane deflection by \(x\), the mass-normalized equation of motion is
\begin{equation} 
\ddot{x}+2\zeta\dot{x}+\omega^2 x+\gamma x^2+\beta x^{3}= F\cos(\omega_\mathrm{d} t),
\label{eqn:0}
 \end{equation}
where \(\zeta\) is the damping ratio and \(\omega\) is the resonance frequency. Furthermore, \(\gamma\) and \(\beta\) are the mass-normalized quadratic and cubic nonlinear coefficients, respectively. The term \(F\cos(\omega_\mathrm{d} t)\) represents the mass-normalized harmonic drive, where \(\omega_\mathrm{d}\) is the drive frequency. For tension-dominated membranes, the resonance frequency is directly related to the membrane tension, such that \(\omega^2\propto n\). In addition, the Duffing coefficient (also known as cubic stiffness term) scales with the Young's modulus, \(\beta\propto E\) \cite{davidovikj2017nonlinear}, whereas the quadratic coefficient scales with the static out-of-plane deflection, \(\gamma\propto x_\mathrm{eq}\) \cite{li2026cascade}. Consequently, the quadratic term vanishes for initially flat membranes. When the membrane is buckled or deflected, however, the quadratic nonlinearity also contributes to the amplitude-dependent frequency shift and can be incorporated into an effective cubic coefficient, \(\beta_\mathrm{eff}=\beta-\frac{10}{9}\frac{\gamma^2}{\omega^2}\), which governs the observed hardening or softening nonlinear responses \cite{Steeneken2021Dynamics2D}. 

For membranes with finite optical absorption, we note that a fraction of the optical energy is converted into heat, resulting in an increase in membrane temperature,
\begin{equation} 
\Delta T = \eta \left( 1+ M \cos{\left[\frac{4\pi x}{\lambda}\right]}\right) P_{\lambda} ,
\label{eqn:1}
\end{equation}
where the temperature rise $\Delta T$ depends not only on the incident laser power, $P_{\lambda}$, but also on the distance of the membrane from the bottom of the cavity (Fig. \ref{fig:fig1}b), which is approximately equal to the cavity depth $d$, through the position-dependent optical intensity. Here, $ \eta $ is the photothermal responsivity of the membrane, $M$ is a factor that depends on the cavity properties and $ \lambda $ is the wavelength of the laser used.  A detailed description of $ \eta$ and $M$ can be found in the Supplementary Information.

The temperature increment $\Delta T$ generates thermally induced stresses in the membrane through thermal expansion. Since $\Delta T$ depends on the membrane position within the optical wave, the resulting restoring force becomes displacement dependent, thereby establishing a photothermal interaction between the optical field and the mechanical motion. The resulting photothermal tension can be decomposed into the sum of a static  and a dynamic contribution ($n=n_\mathrm{s}+n_\mathrm{d}$), in which the static term is determined by the equilibrium position $x_\mathrm{eq}$, while the dynamic term, is determined by the  motion around that equilibrium position. Here, the static component modifies the tension of the membrane at equilibrium and therefore its linear stiffness and resonance frequency, whereas the dynamic component introduces a displacement-dependent restoring force that alters the nonlinear mechanical response. In the limit of small oscillation amplitudes ($\Delta x \ll \lambda/4$), and assuming that only the membrane heats up, as typically encountered in thermally driven motion, the membrane oscillates around $x_\mathrm{eq}$ and the effective membrane static tension becomes
\begin{equation} 
n_\mathrm{s}=n_\mathrm{0}-\frac{Eh\alpha \eta}{1-\nu}
\left(1+M\cos{\left[\frac{4\pi}{\lambda}x_\mathrm{eq}\right]}\right)P_{\lambda} ,
\label{eqn:3}
 \end{equation}
where $h$ is the membrane thickness, $\alpha$ is the effective thermal expansion coefficient, and $\nu$ is the Poisson ratio. Using Eq.~\ref{eqn:3} we calculate the photothermally modified resonance frequency as (see Supplementary Information for the detailed derivation),
\begin{equation} 
\omega^2={\omega_\mathrm{0}}^2 \left[ 1-\frac{Eh\alpha\eta}{n_\mathrm{0} \left(1-\nu\right)}
\left(1+M\cos{\left[\frac{4\pi}{\lambda}x_\mathrm{eq}\right]}\right)P_{\lambda} \right].
\label{eqn:4}
\end{equation}
Equation~\ref{eqn:4} shows that the resonance frequency consists of the intrinsic value $\omega_\mathrm{0}$ and a photothermally induced correction. This correction scales linearly with the optical power and is periodically modulated by the equilibrium position within the cavity. As a result, the resonance frequency can be tuned not only through the incident laser power but also by changing the membrane position within the optical cavity (Fig.~\ref{fig:fig1}c).

For larger oscillation amplitudes, particularly when the membrane is driven into the nonlinear regime, the dynamic component of the photothermal tension ($n_\mathrm{d}$) becomes relevant. Expanding this displacement-dependent contribution around the equilibrium position yields
\begin{equation}  
\begin{split}
n_\mathrm{d}(\Delta x)
=
-{\omega_\mathrm{0}}^2\frac{E h \alpha  \eta}{n_\mathrm{0}(1-\nu)}M
\Bigg(
&-\frac{4\pi}{\lambda}\sin{\left[\frac{4 \pi x_\mathrm{eq}}{\lambda}\right]} \Delta x \\
&-\frac{8\pi^2}{\lambda^2}\cos{\left[\frac{4 \pi x_\mathrm{eq}}{\lambda}\right]} \Delta x^2
+\cdots
\Bigg) P_{\lambda} ,
\end{split}
\label{eqn:5}
\end{equation}
where $\Delta x=x-x_\mathrm{eq}$. Owing to the high thermal conductivity of graphene, we assume that the photothermal response can be treated as quasi-static, such that the membrane temperature follows its displacement without a measurable delay. Here, the displacement-dependent terms generate nonlinear restoring forces. Restricting the analysis to the leading nonlinear contribution, the effective Duffing coefficient becomes
\begin{equation}  
    \beta_\mathrm{eff}=\beta_\mathrm{0}+\frac{8 \pi^2}{\lambda^2} \rho M {\omega_\mathrm{0}}^2 \cos{\left( \frac{4 \pi}{\lambda} x_\mathrm{eq}   \right)}P_{\lambda} - \frac{10}{9} \frac{ {\left(\gamma_\mathrm{0}+\frac{4\pi}{\lambda} \rho M {\omega_\mathrm{0}}^2 \sin{\left[ \frac{4 \pi}{\lambda} x_\mathrm{eq} \right] } P_{\lambda}\right)  }^2}{\omega^2(P_{\lambda},x_\mathrm{eq})}.
    \label{eqn:6}
\end{equation}
where   $ \rho=\frac{Eh\alpha\eta}{n_\mathrm{0}(1-\nu)} $. Equation \ref{eqn:6} shows that the nonlinear stiffness consists of these intrinsic contributions as well as a photothermally induced correction that can be controlled through both the optical power and the equilibrium position. Unlike the resonance frequency shift, the nonlinear correction exhibits a $\pi$-phase offset with respect to the cavity position and can exceed the intrinsic contribution (Fig. \ref{fig:fig1}d).  More specifically, Equation~\ref{eqn:6} reveals two mechanisms through which the nonlinear response can be tuned. First, the direct photothermal correction to the cubic stiffness is proportional to $\cos\!\left(4\pi x_\mathrm{eq}/\lambda\right)$ and therefore changes sign as the equilibrium position is varied within the optical cavity. Thus, for cavity positions satisfying $\cos\!\left(4\pi x_\mathrm{eq}/\lambda\right)<0$, the photothermal contribution increases the effective Duffing coefficient, whereas for $\cos\!\left(4\pi x_\mathrm{eq}/\lambda\right)>0$ it promotes softening behavior (please note that for graphene $\rho<0$). A second contribution arises from the quadratic nonlinearity which always provides a negative correction to $\beta_{\mathrm{eff}}$. The magnitude of this term increases with an increase in the effective quadratic nonlinearity, $\gamma=\gamma_\mathrm{0}+ \frac{4\pi}{\lambda}\rho M{\omega_\mathrm{0}}^2 \sin\left(\frac{4\pi}{\lambda}x_{\mathrm{eq}}  \right)P_{\lambda}$, or by a reduction in the resonance frequency due to photothermal heating. Consequently, buckled or statically deflected membranes, as well as resonators exhibiting strong photothermal softening, display an enhanced tendency toward softening nonlinear dynamic behavior.\\

\begin{figure}[H]
\centering
    \includegraphics[width=0.5\linewidth]{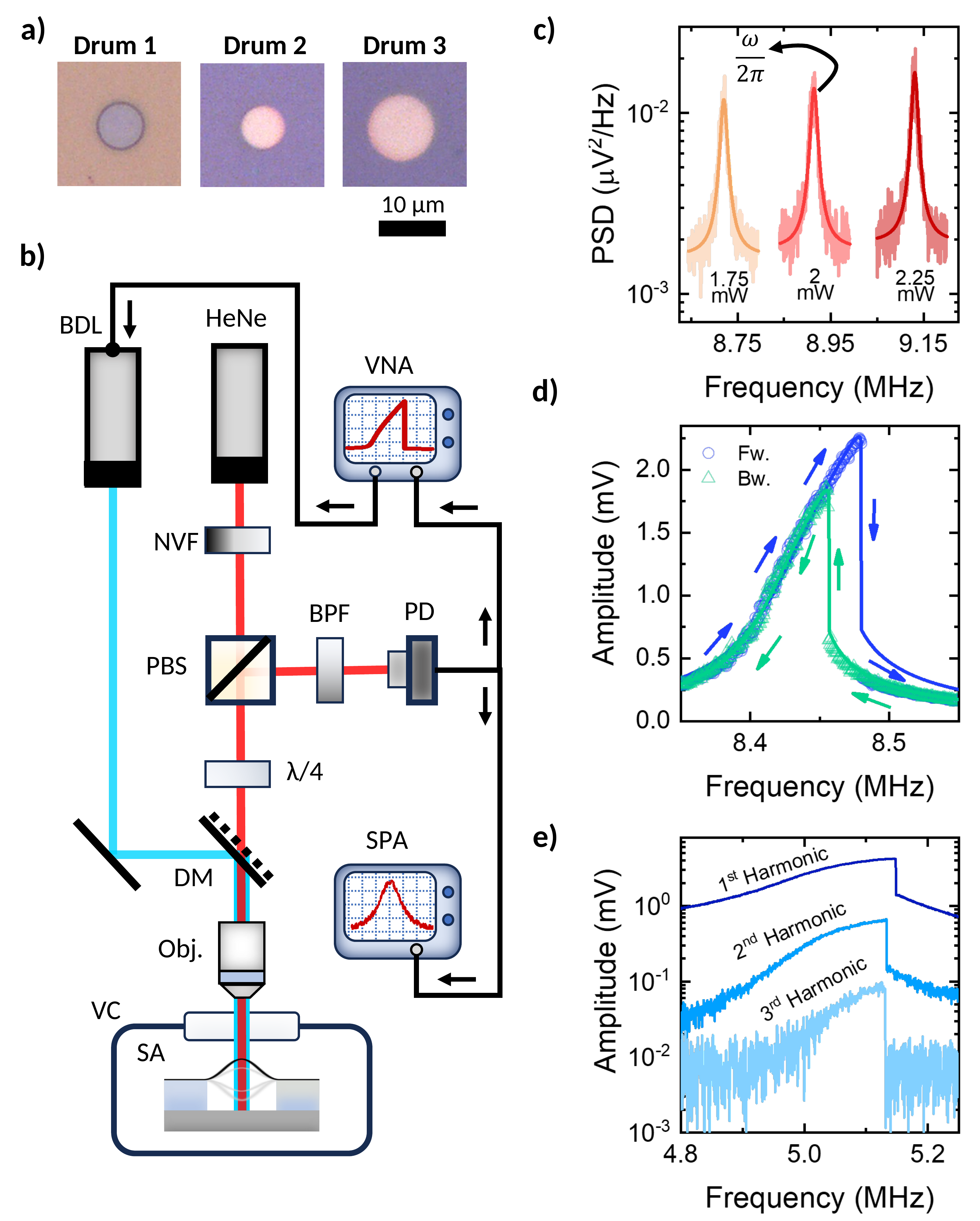}
    \caption{\textbf{Experimental setup.} \textbf{a)} Microscopy image of the devices used. Drum 1: 8 $\mu m$ diameter and 420 nm cavity depth depth. Drum 2: 8 $\mu m$ diameter and 285 nm cavity Drum 3:  12 $\mu m$ diameter and 285 nm cavity depth.  \textbf{b)} Schematic of the setup used in the experiments. Legend: HeNe. Helium-Neon laser. NVF. Neutral variable density filter. PBS. Polarizing beam splitter. BPF. Optical bandpass filter. PD. Photodetector. BDL. Blue diode laser. DM. Dichroic mirror. $\lambda /4)$. $\lambda /4$ waveplate. Obj. Objective. SPA. Spectrum analyzer. VC. Vacuum chamber. SA. Sample  \textbf{c)} Brownian motion spectra acquired for Drum 1 at different red laser powers (PSD: power spectral density). \textbf{d)} Driven mechanical response for Drum 1 showing a nonlinear response. \textbf{e)} First three harmonics measured for Drum 3.}
    \label{fig:fig2}
\end{figure}

To validate the analytical predictions, we measure the mechanical response of circular bilayer graphene membranes suspended over a circular hole in a silicon oxide layer grown on a polished silicon substrate (Fig \ref{fig:fig2}a). The measurements are carried out using a custom-built optical interferometer (Fig.~\ref{fig:fig2}b), where the optical power serves as a control knob to tune the tension and thus $x_\mathrm{eq}$.  A red Helium-Neon laser ($\lambda$ = 632.8 nm) is tightly focused onto the membrane to transduce its motion interferometrically. The reflected signal, whose intensity is modulated by the membrane motion, is detected with an avalanche photodetector and analyzed using a spectrum analyzer (Rohde \& Schwarz FPL1003), allowing the acquisition of the thermomechanical (Brownian) motion spectrum of the resonator under different optical powers (Fig.~\ref{fig:fig2}c). In addition, the setup includes a power-modulated blue diode laser ($\lambda$=488 nm), which serves both as a controllable source of photothermal heating and as an actuator for driving the mechanical motion. The blue laser provides a static optical-power component, that determines the average absorbed optical power, and a modulated component, that generates a periodic photothermal drive force. Since the modulated contribution has zero average value and is substantially smaller than the static component, its effect on the average membrane temperature can be neglected. By synchronizing the modulation signal with the interferometric readout using a vector network analyzer (Rohde \& Schwarz ZNB4), we measure the driven frequency response of the resonators to access the nonlinear Duffing regime (Fig.~\ref{fig:fig2}d). At sufficiently large amplitudes, higher harmonics become visible in the optical response (Fig.~\ref{fig:fig2}e). These harmonics are used to extract both the oscillation amplitude and the equilibrium position $x_\mathrm{eq}$ of the membrane using the calibration method introduced in Ref. \citenum{Dolleman2017AmplitudeCalibration}. A detailed description of the procedure is given in the Supplementary Information.\\

\begin{figure} 
    \centering
    \includegraphics[width=0.5\linewidth]{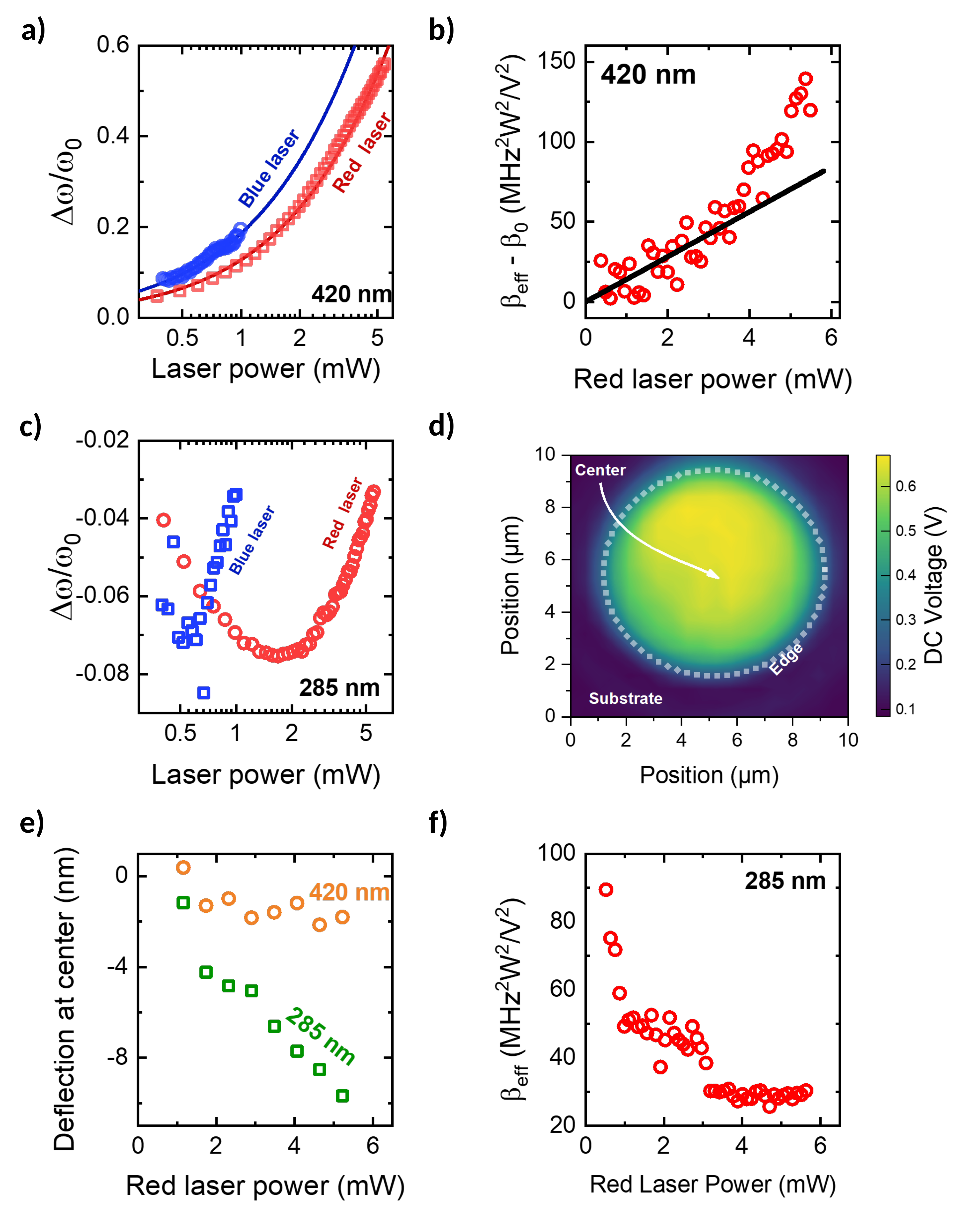}
     \caption{\textbf{Stiffness as a function of power. a)} Resonance frequency shift measured for the 8 $\mu$m diameter membrane suspended over a 420 nm cavity depth (Drum 1) as a function of the power of the red (open circles) and the blue laser (open squares) and their respective fits to the proposed model (solid lines). \textbf{b)}  Effective Duffing parameter measured  for the previous membrane as a function of the red laser power (open circles) and its fit to a line (solid line) \textbf{c)} Resonance frequency shift measured for the 8 $\mu$m diameter membrane suspended over a 285 nm cavity depth (Drum 2) as a function of the power of the red (open circles) and the blue laser (open squares) and their respective fits to the proposed model (solid lines). \textbf{d)} Reflected light intensity mapped in Drum 2. \textbf{e)} Deflection of both membranes measured at their center as a function of the red laser power analyzing the reflected light intensity. \textbf{f)} Effective Duffing parameter measured  for the previous membrane as a function of the red laser power (open circles) for Drum 2.}
    \label{fig:fig3}
\newpage

\end{figure}

To investigate the experimental control over the stiffness as predicted by the analytical model (Eq. \ref{eqn:3} and \ref{eqn:4}) we characterize the dynamics of a 8 $\mu m$ diameter graphene membrane suspended over a 420 nm cavity under the interferometer using different laser powers. To probe the linear stiffness we measure the Brownian motion of the membrane (Fig. \ref{fig:fig3}a), and obtain the resonance frequency by fitting the resulting curve to the harmonic oscillator model (Fig.\ref{fig:fig2}c). Fig. \ref{fig:fig3}a shows that the resonance frequency increases both with red and blue laser power, following the theoretical prediction (Eq. \ref{eqn:4}) with high accuracy ($R^2>0.97$), while assuming a constant value of $x_\mathrm{eq}$. It shall be  noted that the resonance frequency increases with the power since graphene has a negative effective thermal expansion coefficient at room temperature $\alpha<0$. 

After having qualitatively confirmed the linear tuning of the resonance frequency $\omega$, we analyze the effect of photothermal heating on the cubic nonlinear stiffness. To this end, we drive the membranes by modulating the blue laser and measuring the response using a vector network analyzer under different conditions of the static laser power (non-modulated component). Each mechanical spectrum obtained in this case is fitted to a Duffing resonator model (Fig. \ref{fig:fig2}d), obtaining the effective Duffing coefficient $\beta$. Here, it shall be noted that, changing the power of the red laser also changes the transduction factor (i.e. the proportionality constant relating the measured voltage with the membrane displacement). Nevertheless, in these drums the transduction factor cannot be obtained given the signal to noise ratio is not high enough to measure the third harmonic, which impedes a direct comparison of the cubic stiffness for these results. To enable a consistent comparison, we normalize each spectrum to the power of the red laser (directly proportional to the transduction factor) to eventually obtain the cubic  stiffness by fitting. The data obtained for the effective Duffing coefficient (Fig. \ref{fig:fig3}b) reveal that, for lower powers of the red laser, the effective cubic stiffness increases following a linear trend, which agrees well with the analytical model. For higher red laser power ($P> 5$ mW), the change in the effective Duffing constant deviates from the linear trend, which can be explained with the change in the resonance frequency (Fig. \ref{fig:fig3}b).

The analytical model also predicts a negative trend in the shift of the effective cubic stiffness for certain values of the equilibrium position of the membrane. Consequently, to analyze such cases, we measure a second graphene membrane with the same diameter but suspended over a 285 nm cavity. In this case, we measure the Brownian motion for different values of the laser power following the same procedure as in the previous case (Fig. \ref{fig:fig3}c). The resonance frequency of this device shows a non-monotonic dependence on optical power, first decreasing and subsequently increasing as either the red or the blue laser power is raised. This trend is inconsistent with the analytical model, which assumes an initially flat membrane and therefore predicts a monotonic frequency shift. Instead, the observed behavior closely resembles that previously reported for thermally buckled membranes \cite{Liu_2024}. 

To verify the presence of buckling, we therefore measure the equilibrium position of the membranes as a function of optical power for both cavity depths by mapping the reflected intensity in the sample, comparing the reflected intensity in the center of the cavity and outside the cavity (Fig. \ref{fig:fig3}d). A complete description of this procedure is provided in the Supplementary Information.  We notice that the membrane suspended above the 285 nm cavity shows a power-dependent shift of the equilibrium position of up to 10 nm (Fig. \ref{fig:fig3}e), confirming a buckled membrane configuration. In contrast, the membrane suspended above the 420 nm cavity shows no measurable change in its equilibrium position over the investigated power range, consistent with the absence of buckling. Interestingly, for the buckled membrane, the measured effective Duffing coefficient as a function of laser power (Fig.~\ref{fig:fig3}f) decreases with increasing optical power, in agreement with the analytical model. This reduction can be particularly strong when the accompanying reduction in resonance frequency further amplifies the negative contribution to $\beta_{\mathrm{eff}}$ (see Eq.~ \ref{eqn:6}). Although buckling invalidates the assumptions used to derive the resonance frequency model (Eq.~\ref{eqn:4}), the analytical expression for the effective Duffing coefficient (Eq.~\ref{eqn:6}) remains applicable provided that the experimentally measured resonance frequency is used in the evaluation of $\beta_{\mathrm{eff}}$.

To further validate the proposed nonlinear stiffness tuning mechanism and explore its limits, we perform measurements on a third membrane of a larger diameter (12 $\mu m$) suspended over a 285 nm cavity. The larger diameter offers two advantages. First, the increased oscillation amplitude facilitates the detection of higher harmonics, enabling calibration of the displacement using the method proposed by Dolleman \textit{et al.}~\cite{Dolleman2017AmplitudeCalibration}. Second, the lower mechanical stiffness makes the resonator more susceptible to photothermal tuning, thereby enhancing the optomechanical effects predicted by the model. 

\begin{figure}[H]
\centering
\includegraphics[width=\linewidth]{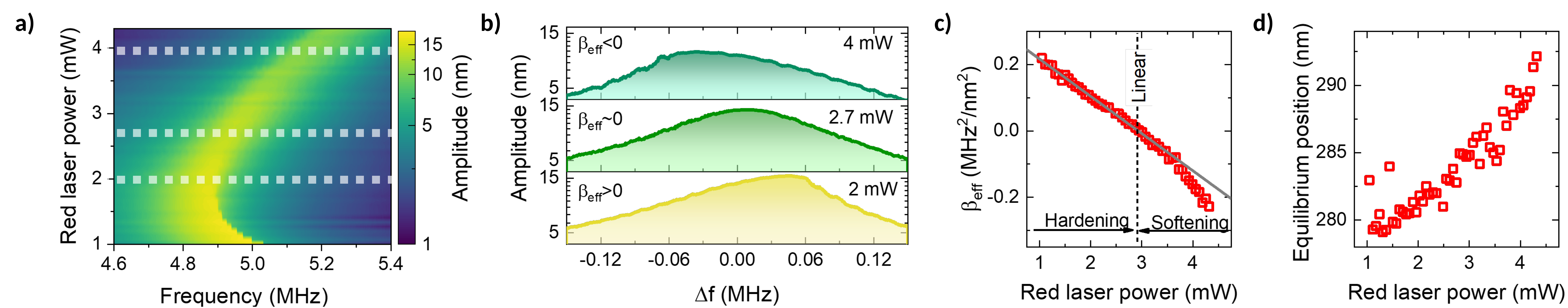}
    \caption{\textbf{Cubic stiffness switching. a)} Mechanical response measured for a 12 $\mu m$ diameter resonator suspended over a 285 nm cavity (Drum 3) as a function of the power of the red laser. White dotted lines represent the data shown in the next plot \textbf{b)} Mechanical response of the membranes for 2 mW, 2.7 mW and 4 mW. \textbf{c)} Cubic stiffness as a function of the power of the red laser obtained from the fitting of the previous data.  \textbf{d)} Equilibrium position of the membrane measured as a function of the power of the red laser.}
    \label{fig:fig4}
\end{figure}

As shown in Fig.~\ref{fig:fig4}a, the resonance frequency exhibits the same non-monotonic dependence on optical power observed for the 8 $\mu$m membrane suspended above a 285 nm cavity (Fig. \ref{fig:fig3}c), confirming that the larger membrane also undergoes photothermally induced buckling. Owing to the enhanced photothermal response, the nonlinear stiffness now undergoes a transition from hardening ($\beta_{\mathrm{eff}}>0$) to softening ($\beta_{\mathrm{eff}}<0$) nonlinear dynamics as the red laser power is increased (Fig.~\ref{fig:fig4}b). At an intermediate power of approximately 2.7 mW, the resonance response becomes linear, indicating that the photothermally induced softening compensates the intrinsic geometric hardening of the membrane, resulting in an almost vanishing effective Duffing coefficient. As in Figs. \ref{fig:fig3}b \& \ref{fig:fig3}f, the effective Duffing coefficient extracted from the nonlinear response curves decreases approximately linearly with increasing optical power (Fig.~\ref{fig:fig4}c), in agreement with the analytical model. At higher powers, however, the measurements deviate from the linear trend. This deviation coincides with the onset of a pronounced shift in the equilibrium position of the membrane (Fig.~\ref{fig:fig4}d), confirming that buckling becomes increasingly important and modifies the resonance frequency, thereby enhancing the photothermal softening predicted by Eq.~\ref{eqn:6}.\\

In summary, we have established both theoretically and experimentally that optical interferometric readout and actuation in suspended 2D material resonators give rise to an optomechanical interaction that directly modifies their mechanical properties through photothermal heating. By controlling the optical power, we achieve active tuning of both the linear stiffness and the cubic nonlinear stiffness of graphene resonators, including continuous transitions between hardening, linear, and softening nonlinear dynamics. Furthermore, we show that photothermal heating can induce or enhance buckling in membranes with low  pretension, providing an additional physical knob for tailoring their nonlinear response. Beyond providing an in situ means to compensate for fabrication-induced variations in membrane tension and nonlinear stiffness, the proposed analytical framework allows intrinsic mechanical properties of graphene resonators to be distinguished from photothermal effects introduced by the optical readout and actuation. More broadly, our results establish optical interferometry not only as a transduction technique but also as a versatile tool for engineering and controlling the dynamics of low-dimensional nanomechanical resonators.

\begin{acknowledgement}
A.M. acknowledges the funding received from the European Union through a postdoctoral fellowship within the framework of the Marie Sklodowska-Curie Actions (MSCA-PF, PROPHOTOM, ID: 101104522). P.G.S. acknowledges funding from the European Union under the Horizon Europe Programme, Grant Agreement No. 101136388. F.A. further acknowledges funding from the European Union (ERC Consolidator, NCANTO, 101125458). Views and opinions expressed are, however, those of the authors only and do not necessarily reflect those of the European Union or the European Research Council. Neither the European Union nor the granting authority can be held responsible for them. The authors also acknowledge SoundCell B.V. for providing the graphene samples for the measurements.
\end{acknowledgement}

\section{Data Availability}
All the experimental data included in this paper are published in the open repository \href{https://data.4tu.nl/}{4TU.ResearchData} under the title "Optomechanical tuning of nonlinearity in graphene resonators" and can be freely accessed and used under a \href{https://creativecommons.org/licenses/by-nc-sa/4.0/deed.en}{CC BY-NC-SA 4.0} license.

\section{Authors contribution}
A.M. conceived and designed the work. A.M. and F.A. developed the analytical model. A.M. performed the experiments and wrote the manuscript with input from all authors. All authors analyzed the data, discussed the results and commented on the manuscript.

\bibliography{references}

\newpage
\newpage

\newpage

\end{document}

\renewcommand{\theequation}{S.\arabic{equation}}
\setcounter{equation}{0}

%\section
  %{Supplementary Information: Optomechanical tuning of nonlinearity in graphene resonators}
  %\begin{center}
     %   \textbf{Alberto Martín-Pérez*, Peter G. Steeneken, Farbod Alijani} \\
        %\textit{Department of Precision and Microsystems Engineering, Faculty of Mechanical Engineering, Delft University of Technology, Netherlands}
        %\\  *Email: A.MartinPerez@tudelft.nl
  %\end{center}

\subsection{S.1. Photothermal heating of the membrane}
The suspended membrane and the reflective silicon substrate used in the experiments form an optical cavity so when the lasers of the optical setup are focused on this cavity, interference is produced between the incident beam and the different reflected and transmitted beams. Consequently, the intensity ($\mathrm{I}$) at the membrane surface it is the sum of the intensities of the different beams as well as their relative phase difference (Eq. \ref{eqn:s1}).
\begin{equation}
    I=I_\mathrm{inc}+I_\mathrm{ref}+2\sqrt{I_\mathrm{inc}I_\mathrm{ref}} \cos{(\Delta \varphi )}
    \label{eqn:s1}
\end{equation}
Here  $I_\mathrm{inc}$ is the incident intensity of the beam, $I_\mathrm{ref}$ the reflected intensity and $\Delta \varphi$ the phase difference among these beams. 

Considering the different layers of the cavity (2D material/vacuum gap/ideal mirror), the reflected intensity will result as the product of the reflectivity, transmission and absorption of the different layers. In the experiments the light used has polarized circularly, as a result this interference can be analyzed by decomposing the light into two orthogonal components (P and S) with the same intensity.
\begin{equation} \begin{split} \begin{aligned}
    I_\mathrm{inc,P}&=\frac{1}{2}I_0 (1-R_\mathrm{vg,p})e^{(-\varepsilon \Delta)} -\frac{1}{2}I_0(1-R_\mathrm{gv,p})R_\mathrm{vg,p}e^{(-2\varepsilon \Delta)} 
\\    
 I_\mathrm{inc,S}&=\frac{1}{2}I_0 (1-R_\mathrm{vg,s})e^{(-\varepsilon \Delta)} +\frac{1}{2}I_0(1-R_\mathrm{gv,s})R_\mathrm{vg,s}e^{(-2\varepsilon \Delta)} 
  \label{eqn:s2}
 \end{aligned}\end{split} \end{equation}
\begin{equation} \begin{split} \begin{aligned}
I_\mathrm{ref,P} &=\frac{1}{2}I_0\left(1-R_\mathrm{vg,p}\right)^3e^{-2 \varepsilon \Delta}  							
\\
I_\mathrm{ref,S} &=\frac{1}{2}I_0\left(1-R_\mathrm{vg,s}\right)^3e^{-2 \varepsilon \Delta}  
\end{aligned} \end{split}
  \label{eqn:s3}
 \end{equation}
where $ R_\mathrm{vr,p}$ and $ R_\mathrm{vr,s}$  are the reflectivity in the vacuum-resonator interphase for the P and S polarizations respectively, $ R_\mathrm{rv,p}$ and $R_\mathrm{rv,s}$ the reflectivity in the interphase resonator-vacuum for the P and S polarizations respectively, $\varepsilon$ the extinction coefficient in the resonator and $\Delta$ the thickness of the resonator.

For the sake of simplicity, we consider only the first reflections in the cavity, so the terms $R_\mathrm{gv}$ and $R_\mathrm{vg}$ can be obtained directly from the Fresnel equations. Moreover, the experimental conditions impose the normal incidence of the beam on the cavity, consequently, the value of the reflectivity will be the same for the P and S components of the polarization while $ R_\mathrm{gv}$ and $R_\mathrm{vg}$ terms can be simplified as,
\begin{equation} 
R_\mathrm{rv,s}=R_\mathrm{rv,p}=R_\mathrm{rv}=\left(\frac{n_\mathrm{r}-n_\mathrm{m}}{n_\mathrm{r}+n_\mathrm{m}}\right)^2
\label{eqn:s6}
 \end{equation}
\begin{equation} 
R_\mathrm{vr,s}=R_\mathrm{vr,p}=R_\mathrm{vr}=\left(\frac{n_\mathrm{m}-n_\mathrm{r}}{n_\mathrm{r}+n_\mathrm{m}}\right)^2
\label{eqn:s7}
\end{equation}
Where $ n_\mathrm{r}$ is the refractive index in the resonator and $ n_\mathrm{m}$ the refractive index of the surrounding medium. It shall be noted that here the surrounding medium is always vacuum ($n_\mathrm{m}=1$), however, in section S.3. we also consider that the surrounding medium can be the silicon oxide layer that separates the graphene membrane from the reflective substrate ($n_\mathrm{m}=n_\mathrm{ox}$). The previous expressions yield $R_\mathrm{rv}=R_\mathrm{vr}$, allowing to simplify the intensity reflected by the cavity.

Furthermore, we need to consider the phase between the incident and reflected beam. Here, it shall be noted that since we are using circular polarized light, a $\pi$ phase is introduced in the S component of the reflected beams (but not for the P component). Given the experimental conditions, there is normal incidence in the cavity, allowing the phase difference to be written as follows
\begin{equation} \begin{split} \begin{aligned}
\Delta \varphi_\mathrm{S} &=\pi+\frac{4 \pi}{\lambda}(n_\mathrm{r} \Delta + n_\mathrm{vac}d)
\\
\Delta \varphi_\mathrm{P} &=\frac{4 \pi}{\lambda}(n_\mathrm{r} \Delta + n_\mathrm{vac}d)
\end{aligned}\end{split} \label{eqn:s9}
 \end{equation}
Here, $\lambda$ is the laser wavelength, $n_\mathrm{r}$ is the refractive index of the resonator, and $n_{\mathrm{vac}}$ is the refractive index of vacuum. Because the membrane thickness is much smaller than the vacuum-gap depth, $\Delta \ll d$, and $n_{\mathrm{vac}} \approx 1$, the optical path length can be approximated as\[n_\mathrm{r} \Delta + n_{\mathrm{vac}} d \approx d.\]
Substituting the previous expressions in Eq. \ref{eqn:s1}, we obtain the intensity as a result of the interference 
\begin{equation} \begin{split} \ \begin{aligned}
I_\mathrm{P}&=\frac{1}{2}I_\mathrm{0}R_\mathrm{0,P}\left[1+M_\mathrm{P}\cos{\left(\frac{4\pi}{\lambda}n_\mathrm{vac}d\right)}\right]
\\
I_\mathrm{S}&=\frac{1}{2}I_\mathrm{0}R_\mathrm{0,S}\left[1-M_\mathrm{S}\cos{\left(\frac{4\pi}{\lambda}n_\mathrm{vac}d\right)}\right]  
\end{aligned}\end{split}
\label{eqn:s11}
\end{equation}
Where $M$ and $R_\mathrm{0}$ are the modulation and reflectance factors respectively, defined as,
\begin{equation} \begin{split} \begin{aligned}
R_\mathrm{0,P} &=(1-R_\mathrm{vg}e^{-\varepsilon \Delta}+[1-R_\mathrm{vg}]^2e^{-\varepsilon \Delta})(1-R_\mathrm{vg})e^{-\varepsilon \Delta} 	
\\
R_\mathrm{0,S} &=(1+R_\mathrm{vg}e^{-\varepsilon \Delta}+[1-R_\mathrm{vg}]^2e^{-\varepsilon \Delta})(1-R_\mathrm{vg})e^{-\varepsilon \Delta} 
\end{aligned}\end{split}
\label{eqn:s13}
\end{equation}
\begin{equation} \begin{split} \begin{aligned}
M_\mathrm{P} &=\frac{2(1-R_\mathrm{vg}) \sqrt{(1-R_\mathrm{vg}e^{-\varepsilon \Delta})e^{-\varepsilon \Delta}}}{R_\mathrm{0,P}}  
\\
M_S &=\frac{2(1-R_\mathrm{vg}) \sqrt{(1+R_\mathrm{vg}e^{-\varepsilon \Delta})e^{-\varepsilon \Delta}}}{R_{0,S}} .	
\end{aligned} \end{split}
\label{eqn:s15}
\end{equation}
To calculate the photothermal heating produced in the membrane, we can treat the intensity as the sum of its S and P components assuming the material is homogeneous and isotropic
\begin{equation} 
I=I_\mathrm{0}R_\mathrm{0}\left[1+M\cos{\left(\frac{4\pi}{\lambda}d\right)}\right] 	
\label{eqn:s16a}
\end{equation}
With
\begin{equation}
R_\mathrm{0}=\frac{1}{2}R_\mathrm{0,P}+\frac{1}{2}R_\mathrm{0,S} 
\label{eqn:s16}
 \end{equation}
\begin{equation} 
M=\frac{\frac{1}{2}R_\mathrm{0,P}M_P-\frac{1}{2}R_\mathrm{0,S}M_S}{\frac{1}{2}R_\mathrm{0,P}+\frac{1}{2}R_\mathrm{0,S}} 	
\label{eqn:s17}
\end{equation}
Here the factors $\frac{1}{2}$ are due to the circular polarization of light, however, for other polarization states, the previous expressions can be used after replacing these fractions with those corresponding to the new polarization.

Next, using the light intensity at the membrane given by Eq.~\ref{eqn:s16a}, we calculate the temperature increase caused by optical absorption. Only a fraction of the incident light is absorbed by the membrane; this fraction is described by the wavelength-dependent absorbance, $A_{\lambda}$. The absorbed power is therefore obtained by multiplying the absorbance by the light intensity integrated over the illuminated area. For simplicity, we approximate this area as a circle over which the light intensity is uniform
\begin{equation}
 P_\mathrm{abs}=A_{\lambda} I\pi{r_\mathrm{s}}^2 
\label{eqn:s18}
\end{equation}
where $r_\mathrm{s}$ the radius of the laser spot when it is focused on the membrane.
Assuming the heat transmission in the membrane is produced exclusively by conduction, we calculate the temperature increment by solving the stationary heat equation. To solve the heat equation we make a number of assumptions for simplicity. First, given the geometry of the problem, we can assume that the heat will be transmitted from the point where the laser is focused to the rest of the membrane in concentric circles. This cylindrical symmetry ($ {\frac{\partial^2T}{\partial\theta^2}=0}$) reduces the problem to the heat transmission in the radial direction. Moreover, we can divide the space in two regions: inside the laser spot (where the gradient of the heat flux vanishes) and outside the laser spot (where the heat is transmitted in concentric circles). With these assumptions, we obtain the following piecewise differential equation
\begin{equation} 
\begin{cases}
   % \begin{multline*}  
    %\mathrm{

\kappa\left(\frac{\partial^2T}{\partial r^2}+\frac{1}{r}\frac{\partial T}{\partial r}\right)=0\ \ \qquad r\le r_\mathrm{s} \\
\kappa\left(\frac{\partial^2T}{\partial r^2}+\frac{1}{r}\frac{\partial T}{\partial r}\right)=\frac{2P_\mathrm{abs}}{\pi h r^3}\ \ \ \ r>r_\mathrm{s}
% }
   % \end{multline*}
\label{eqn:s19}
\end{cases}
\end{equation}
Solving the previous differential equation and imposing as a boundary condition $T(r\rightarrow \infty)=T_\mathrm{0}$, where $ \mathrm{T_0}$ is the room temperature, then we obtain the following temperature profile.
\begin{equation}
T(r)= \begin{cases} 
\frac{2P_\mathrm{abs}}{\pi\kappa h R_\mathrm{s}}+T_0 \ \ r\le r_\mathrm{s}
\\
\frac{2P_\mathrm{abs}}{\pi\kappa hr}+T_\mathrm{0}
 \ \ \ r>r_\mathrm{s}\ 
\label{eqn:s20}
\end{cases} 
\end{equation}
Given graphene’s high thermal conductivity, the temperature decreases rapidly outside the laser-illuminated region. We therefore approximate the temperature increase as uniform within the laser spot and negligible outside it and thus
\begin{equation}
\Delta T=\frac{2A_{\lambda}}{\kappa h}r_{\mathrm{s}}I_{\mathrm{0}}R_{\mathrm{0}}\left(1+M \cos\left[\frac{4\pi d}{\lambda} \right] \right)P_{{\lambda}}
    \label{eqn:s21}
\end{equation}
For the sake of simplicity, we introduce the photothermal responsivity ($ \eta_\lambda$) so as to have a linear relation between the absorbed intensity and the temperature increment
\begin{equation}
\eta_{\lambda}=\frac{2A_{\lambda}}{\kappa h}r_\mathrm{s}I_\mathrm{0}R_\mathrm{0} 
\label{eqn:s22}
\end{equation}
For graphene, the difference in absorbance between the two wavelengths used in the experiments, 632.8~nm and 488~nm, is negligible \cite{Bonaccorso2010GraphenePhotonics}. We therefore treat the absorbance as wavelength independent in the main text, such that $\eta_{\lambda} = \eta$.

Please note that, for convenience, in this section we used the intensity of light ($I_\mathrm{0}$) instead of the laser power used in the main text. The equations shown in this section can be easily adapted to work with the laser power substituting the intensity as the laser power divided by the spot area.

\subsection{S.2. Equation of motion of the membrane}
The flexural oscillations in suspended graphene are usually modeled as thin membranes in which the pretension is the effect dominating the restoring force. This assumption allows the equation of motion to be approximated as a harmonic oscillator with an effective mass and stiffness. The latter being directly proportional to the pretension of the membrane.
\begin{equation}
k_\mathrm{1}=n\pi\int_{0}^{R}{\left(\frac{\partial\phi(r)}{\partial r}\right)^2rdr}=S n 
\label{eqn:s23}
\end{equation}
where $ k_\mathrm{1}$ is the linear stiffness, $ n$ is the pretension of the membrane and $ \phi$ is the mode shape. For the sake of simplicity, we introduce the geometrical factor $S$ and use it as a proportionality constant for the membrane tension.
Using the effective linear stiffness, we describe the membrane dynamics through the equation of motion of a damped harmonically driven oscillator. To account for geometric nonlinearities in the membrane, we also include quadratic and cubic stiffness terms. Therefore,
\begin{equation}
\ddot{x}+2\zeta\dot{x}+\frac{k_\mathrm{1} (x,P_{\lambda})}{m_\mathrm{eff}}x+\gamma_\mathrm{0}x^2+\beta_\mathrm{0}x^3=F\cos{\left (\omega_\mathrm{d} t \right)}
\label{eqn:s24}
\end{equation}
Here, $m_{\mathrm{eff}}$ is the effective mass, $\zeta$ is the damping ratio, $\gamma_\mathrm{0}$ is the quadratic stiffness coefficient, $\beta_\mathrm{0}$ is the Duffing coefficient, $F$ is the mass-normalized driving-force amplitude, and $\omega_\mathrm{d}$ is the driving frequency. Photothermal heating modifies the membrane pretension and, consequently, its linear stiffness. Both therefore depend on the incident laser power and on the membrane position relative to the reflective substrate, leading to a modified equation of motion. Because the photo-induced tension depends on the membrane displacement, as described by Eq. 3, we expand it in a Taylor series about the equilibrium position, $x_{\mathrm{eq}}$. For simplicity, we retain terms only up to second order in the expansion, which allows the resulting equation of motion to be expressed in the form of a Duffing oscillator.
\begin{equation} \begin{split}
    n = n_\mathrm{0} -\frac{Eh\alpha\eta_\lambda}{1-\nu}
    \Bigg( 1+M\Big[\cos\!\Big( \frac{4\pi}{\lambda}  x_{\mathrm{eq}}\Big)-\frac{4\pi }{\lambda}\sin\!\Big( \frac{4\pi}{\lambda}  x_{\mathrm{eq}}\Big)\Delta x \\ -\frac{1}{2} \Big( \frac{4\pi }{\lambda} \Big)^2 \cos\!\Big( \frac{4\pi}{\lambda}  x_{\mathrm{eq}}\Big) \Delta x^2
+...\Big] \Bigg) P_\lambda
\end{split} \label{eqn:s26} \end{equation}

By substituting the position-dependent expression of the membrane tension (Eq. \ref{eqn:s26}) in the equation of motion and rearranging the terms, we obtain the following equation of motion.
\begin{equation}
    \ddot{x}+2\zeta\dot{x}+{\omega}^2x+\gamma x^2+\beta x^3=F\cos{\omega_\mathrm{d} t} 
    \label{eqn:eom}
\end{equation}
The linear, quadratic and cubic terms in Eq.\ref{eqn:eom} depend on the power of the laser as well as on the equilibrium position of the membrane:
\begin{equation}
{\omega}^2={\omega_\mathrm{0}}^2\left[1-\rho\left(1+M\cos{\left[\frac{4\pi}{\lambda}x_\mathrm{eq}\right]}\right)P_{\lambda}\right ]+\Delta\omega^2_\mathrm{initial}
    \label{eqn:s28a}
\end{equation}
\begin{equation}
\gamma=\gamma_\mathrm{0}+\frac{4\pi}{\lambda}\rho M{\omega_\mathrm{0}}^2\sin{\left[\frac{4\pi}{\lambda}x_\mathrm{eq}\right]}P_{\lambda}  	
    \label{eqn:s28c}
\end{equation}
\begin{equation}
\beta=\beta_\mathrm{0}+\frac{1}{2}\left(\frac{4\pi}{\lambda}\right)^2\rho M {\omega_\mathrm{0}}^2\cos{\left[\frac{4\pi}{\lambda}x_\mathrm{eq}\right]}P_{\lambda} 	
    \label{eqn:s28d}
\end{equation}
With,
\begin{equation}
{\omega_\mathrm{0}}^2=\frac{S n_\mathrm{0}}{m_\mathrm{eff}}    
    \label{eqn:s28b}
\end{equation}
\begin{equation} \rho=\frac{Eh\alpha\eta_{\lambda}}{n_\mathrm{0}(1-\nu)} 
   \label{eqn:s28e}
\end{equation}
For the linear term, we also account for a possible initial frequency shift, $\Delta\omega_{\mathrm{initial}}^2$, to better represent the experimental conditions. In the experiments, two laser wavelengths are applied simultaneously, and each contributes independently to the photothermal heating of the membrane. However, because the power of one laser is varied while that of the other is kept constant, the effect of the constant-power laser can be incorporated into $\Delta\omega_{\mathrm{initial}}^2$ as an offset. Similar offset contributions may also be included in the coefficients $\gamma_\mathrm{0}$ and $\beta_\mathrm{0}$. Finally, because graphene has a negative thermal expansion coefficient, $\alpha < 0$, it follows that $\rho < 0$.

The nonlinear response eventually can be fitted using the following frequency response function
\begin{equation}
    \left[\left( {\omega_\mathrm{d}}^2-\omega^2 -\frac{3}{4} \beta_\mathrm{eff}\Delta x^2\right)^2 + \left(  2\zeta \omega_\mathrm{d}\right)^2 \right]\Delta x^2=F^2    
\label{eqn:s27a}
\end{equation}
With
\begin{equation}
\beta_\mathrm{eff}=\beta(P_{\lambda},x_\mathrm{eq})- \frac{10}{9} \left(  \frac{\gamma(P_\mathrm{lambda,x_\mathrm{eq}})}{  \omega(P_{\lambda},x_\mathrm{eq})} \right)^2  
    \label{eqn:betaeffgral}
\end{equation}
Substituting the analytical values of resonance frequency (Eq. \ref{eqn:s28a}), the quadratic and the cubic terms (Eq. \ref{eqn:s28c} and \ref{eqn:s28d}) in Eq. \ref{eqn:betaeffgral} we obtain the following expression.
\begin{equation}
\begin{split}    \beta_\mathrm{eff}=\beta_\mathrm{0}+\frac{1}{2}\left(\frac{4\pi}{\lambda}\right)^2\rho M {\omega_\mathrm{0}}^2\cos{\left[\frac{4\pi}{\lambda}x_\mathrm{eq}\right]}P_{\lambda}  \\ - \frac{10}{9}
\frac{\left( \gamma_\mathrm{0}+\frac{4\pi}{\lambda}\rho M{\omega_\mathrm{0}}^2\sin{\left[\frac{4\pi}{\lambda}x_\mathrm{eq}\right]}P_{\lambda} \right)^2}
{{\omega_\mathrm{0}}^2\left[1-\rho\left(1+M\cos{\left[\frac{4\pi}{\lambda}x_\mathrm{eq}\right]}\right)P_{\lambda}\right ]+\Delta\omega^2_\mathrm{initial}}  \end{split}    
    \label{eqn:s27}
\end{equation}
Moreover, it can be demonstrated that, when $\beta_\mathrm{eff}\Delta x^2 \rightarrow0$ (a condition that is satisfied in the experiments when measuring the Brownian motion) the curve describing the dynamic response of the membrane (Eq. \ref{eqn:s27a}) converges to the harmonic oscillator:
\begin{equation} 
\Delta x (\omega_\mathrm{d}) = \frac{F}{ \sqrt{\left( \omega^2-{\omega_\mathrm{d}}^2 \right) ^2+ \left(2 \zeta \omega_\mathrm{d} \right)^2} }     
\end{equation}

\subsection{S.3. Interferometric measurements of membrane deflection}
In the main text, only the modulated (AC) component of the photodetector signal is used to characterize the membrane dynamics. The non-modulated (DC) component, however, also contains information about the static separation between the membrane and the reflective substrate. This makes it possible to assess features such as membrane flatness or buckling by scanning the reflected intensity across the device. To extract this information, we first obtain the optical intensity collected by the photodetector. This intensity results from the interference of the light reflected after propagating through the different layers of the optical cavity in both directions. Following an analysis analogous to that presented in Section~S.1, the intensity measured at the photodetector can also be described by Eq.~\ref{eqn:s16a}, but with different effective cavity-reflectance and modulation factors. These factors differ because, in the present case, the reflected beam crosses the graphene--vacuum interface once more before reaching the photodetector. In addition, Section~S.1 considered only the suspended region of the membrane, whereas here we also account for the supported regions, where one of the vacuum layers is replaced by silicon dioxide. Taking these differences into account, the corresponding modulation and reflectance factors are the following,
 \begin{equation} \begin{aligned} 
   R_\mathrm{P,suspended} &={1+R_\mathrm{mv}+(1-R_\mathrm{mv})^4} 
\\
    R_\mathrm{S,suspended} &={1-R_\mathrm{mv}+(1-R_\mathrm{mv})^4} 
 \end{aligned} \end{equation}
 \begin{equation} \begin{aligned} 
    M_\mathrm{P,suspended} &=\frac{2\sqrt{(1+R_\mathrm{mv})(1-R_\mathrm{mv})^4}}{R_\mathrm{P,suspended}} 
\\
    M_\mathrm{S,suspended} &=\frac{2\sqrt{(1-R_\mathrm{mv})^5}}{R_\mathrm{S,suspended}}
 \end{aligned} \end{equation}
\begin{equation} \begin{aligned} 
    R_\mathrm{P,non-suspended} &={1+R_\mathrm{mv}+(1-R_\mathrm{mv})^2(1-R_\mathrm{ms})^2} 
    \\
    R_\mathrm{S,non-suspended} &={1-R_\mathrm{mv}+(1-R_\mathrm{mv})^2(1-R_\mathrm{ms})^2}
 \end{aligned} \end{equation}
\begin{equation} \begin{aligned} 
    M_\mathrm{P,non-suspended} &=\frac{2\sqrt{(1+R_\mathrm{mv})(1-R_\mathrm{mv})^2(1-R_\mathrm{ms})^2}}{R_\mathrm{P,non-suspended}} 
    \\
    M_\mathrm{S,non-suspended} &=\frac{2\sqrt{(1-R_\mathrm{mv})^3(1-R_\mathrm{ms})^2}}{R_\mathrm{S,non-suspended}} 
 \end{aligned} \end{equation}
where $R_\mathrm{mv}$ is the reflectivity of the interlayer graphene-vacuum and $R_\mathrm{ms}$ is the reflectivity of the interlayer graphene-substrate.

The intensity of the light source ($I_\mathrm{0}$) could be also calculated analytically considering the power of the laser and the spot size, however, relying on the calculated value of the spot of the laser is not sufficiently accurate and we use an alternative method to obtain this parameter. In the samples used, there are zones in which the membrane is not suspended and, consequently, the distance between the membrane and the substrate is known with good precision. By scanning the sample (Fig. 3d) we can therefore determine the mean intensity in those regions where the membrane is not suspended and with that, determine the value of $I_0$ by substituting in Eq. \ref{eqn:s16} using the nominal values of the modulation factors. Having mapped the reflected intensity in the sample and after calibrating $I_\mathrm{0}$,  we then obtain the relative position of the graphene membrane with respect to the reflective substrate as; 
\begin{equation} 
   x_\mathrm{eq}(y,z)=d_0+\frac{\lambda}{4\pi M_\mathrm{mem}} \left[ \frac{I(y,z)}{I_\mathrm{0}-1}\right] ,
\end{equation}
with
\begin{equation}
    M_\mathrm{mem}=\frac{\frac{1}{2}R_\mathrm{P,suspended}M_\mathrm{P,suspended}-\frac{1}{2}R_\mathrm{S,suspended}M_\mathrm{S,suspended}}{\frac{1}{2}R_\mathrm{P,suspended}+\frac{1}{2}R_\mathrm{S,suspended}} .
\end{equation}
We use this procedure for both membranes using different values of the power of the red laser to determine the equilibrium position of the membrane as a function of the laser power (Fig.3e). 

\subsection{S.4.  Determining the transduction factor}
The method introduced by Dolleman et al \cite{Dolleman2017AmplitudeCalibration} for obtaining the transduction factor and the membrane deflection, relies on expanding the light intensity measured at the photodetector around the equilibrium position by using a Taylor series (Eq. \ref{eqn:s4-1}).
\begin{equation} \ \begin{split} 
I=A+B\cos{\left[\mu x_\mathrm{eq}\right]}
    -B\mu\sin{\left[\mu x_\mathrm{eq}\right]}\Delta x
    -\frac{B\mu^2}{2}\cos{\left[\mu x_\mathrm{eq}\right]}\Delta x^2  \\
    +\frac{B\mu^3}{6}\sin{\left[\mu x_\mathrm{eq}\right]}\Delta x^3
    +\frac{B\mu^4}{24}\cos{\left[\mu x_\mathrm{eq}\right]}\Delta x^4
    + ...   \end{split}
        \label{eqn:s4-1}
\end{equation}
Where A and B are optical parameters that depend on the intensity of the incident beam and the refractive index of the different layers of the cavity (it can be demonstrated that these parameters are composed of the modulation factor [$M$] and total reflectivity [$R_0$] introduced in S.1 and S.3) and $\mu=\frac{4\pi}{\lambda}$.
In Ref.~\cite{Dolleman2017AmplitudeCalibration}, the in-phase quadrature measured with a vector network analyzer is substituted into the preceding expression. The resulting harmonic amplitudes are then used to determine the transduction factor and the equilibrium position. In our measurements, however, the quadrature components of the second and third harmonics have a low signal-to-noise ratio, which prevents a reliable application of the original method. The signal magnitude, by contrast, exhibits a sufficiently high signal-to-noise ratio for the first three harmonics. We therefore adapt the method to use the magnitude of the measured signal rather than an individual quadrature component. Writing the displacement as $\Delta x = a_0 e^{i\theta}$, the preceding equation becomes
\begin{equation}
\begin{aligned}
I ={}& A + B\cos\left(\mu x_{\mathrm{eq}}\right)
-B\mu\sin\left(\mu x_{\mathrm{eq}}\right)a_0 e^{i\theta} \\
&-\frac{B\mu^2}{2}\cos\left(\mu x_{\mathrm{eq}}\right)a_0^2 e^{2i\theta}
+\frac{B\mu^3}{6}\sin\left(\mu x_{\mathrm{eq}}\right)a_0^3 e^{3i\theta} \\
&+\frac{B\mu^4}{24}\cos\left(\mu x_{\mathrm{eq}}\right)a_0^4 e^{4i\theta}
+\cdots .
\end{aligned}
\end{equation}
The magnitudes of the different harmonic components can be obtained from the preceding expression as
\begin{equation}
\begin{aligned}
\left|I_{1\omega}\right|
    &= B\mu
    \left|\sin\left(\mu x_{\mathrm{eq}}\right)\right|a_0, \\
\left|I_{2\omega}\right|
    &= \frac{B\mu^2}{2}
    \left|\cos\left(\mu x_{\mathrm{eq}}\right)\right|a_0^2, \\
\left|I_{3\omega}\right|
    &= \frac{B\mu^3}{6}
    \left|\sin\left(\mu x_{\mathrm{eq}}\right)\right|a_0^3, \\
\left|I_{4\omega}\right|
    &= \frac{B\mu^4}{24}
    \left|\cos\left(\mu x_{\mathrm{eq}}\right)\right|a_0^4.
\end{aligned}
\label{eq:harmonic_magnitudes}
\end{equation}

The membrane displacement amplitude can then be determined from the ratios between harmonic components:
\begin{equation}
a_0=\frac{1}{\mu}
\sqrt{
6\frac{\left|I_{3\omega}\right|}
        {\left|I_{1\omega}\right|}
}=\frac{1}{\mu}
\sqrt{
12\frac{\left|I_{4\omega}\right|}
         {\left|I_{2\omega}\right|}
}.
\label{eq:displacement_from_harmonics}
\end{equation}
The equilibrium position can be obtained in a similar manner. Because only the magnitudes of the harmonic components are measured, the sign of
$\tan\left(\mu x_{\mathrm{eq}}\right)$ cannot be determined. The resulting set of possible equilibrium positions is therefore
\begin{equation}
x_{\mathrm{eq}}
= \frac{1}{\mu}
\left[
n\pi
\pm
\arctan\left(
\frac{\mu a_0}{2}
\frac{\left|I_{1\omega}\right|}
     {\left|I_{2\omega}\right|}
\right)
\right],
\qquad n\in\mathbb{N}.
\label{eq:equilibrium_position}
\end{equation}
\bibliography{references}